\documentclass[12pt]{article}
\usepackage{epsfig,amssymb}
\usepackage{color}
\usepackage{graphicx,color}
\usepackage{epstopdf}
\usepackage{float}

\def\appendix{{\newpage\section*{Appendix}}\let\appendix\section%
        {\setcounter{section}{0}
        \gdef\thesection{\Alph{section}}}\section}

\newcommand{\be}{\begin{equation}}
\newcommand{\ee}{\end{equation}}
\newcommand{\bear}{\begin{eqnarray}}
\newcommand{\eear}{\end{eqnarray}}
\newcommand{\ba}{\begin{array}}
\newcommand{\ea}{\end{array}}

\begin{document}

\begin{flushright}
{\tt hep-th/yymmnnn}
\end{flushright}
\vspace{5mm}

\begin{center}
{{\Large \bf The Thermodynamic Properties of Warped Taub-NUT AdS Black String}\\[14mm]
{Chong Oh Lee}\\[2.5mm]
{\it Department of Physics, Kunsan National University,\\
Kunsan 573-701, Korea}\\
{\tt cohlee@kunsan.ac.kr}}
\end{center}

\vspace{10mm}

\begin{abstract}
When we consider five-dimensional warped Taub-NUT/Bolt AdS black
string with minimally coupled massive scalar field, we calculate
entropy by using the brick wall method. It is found that they are proportional to being
quadratically divergent in a cutoff parameter. In particular, we
show that the entropy of warped Taub-NUT AdS black
string holds for an area law in the bulk as well as on the brane.
Furthermore, when the negative cosmological constant is treated as thermodynamic pressure,
we calculate the thermodynamic quantities and investigate their extended thermodynamic properties.
Interestingly, we obtain a thermodynamically stable range
as a function of the temperature for warped Taub-NUT AdS black string.
Finally, we also study a proportional behavior of the thermodynamic quantities
along a warp factor, and find that an entropy, a specific heat, a Gibbs free energy,
and an action difference increase as a warp factor grows up.
\end{abstract}
\newpage

\setcounter{equation}{0}
\section{Introduction}
One can simply obtain $p$-brane solutions by adding a number of $p$ extra dimensions to solutions
for Einstein vacuum equations in $d$-dimensional spacetime~\cite{Gregory:1993vy,Gregory:1994bj}.
The simplest case of them can be given as black string solutions by adding just one extra dimension
in higher dimensional spacetime.
However, one cannot find asymptotically AdS black string solutions through an trivial way in as the above
asymptotically flat case. They are able to be found after imposing warped geometry or working numerically~
\cite{Copsey:2006br,Mann:2006yi} obtained by solving Einstein equation with an extra dimension.
It is of interest to find the black string solution in Taub-NUT AdS background.

The brick wall model was suggested in~\cite{'tHooft:1984re} search
for black hole with a microscopic quantum states, and as the
explanation of the origin of black hole entropy. It was investigated
through the extensive application of various black holes
~\cite{Mann:1990fk}$-$\cite{Gupta:2013ata}.

In fact, black string forms when the matter field in five-dimensional spacetime is
trapped on four-dimensional spacetime and
undergoes gravitational collapse towards black hole.
Comparing with horizon topology $S^2$ in four-dimensional black hole solution,
its horizon topology is $S^2\times S^1$
since such horizon extends into the extra dimension.
Furthermore, the Bekenstein-Hawking entropy
with the geometric character comes from the contributions of fields near the horizon.
In this context, it was probed with black string~\cite{Liu:2008zzr}.
In particular, it is shown that  the entropy in Taub-NUT metric is quadratically
divergent in a cutoff parameter~\cite{Ghosh:2002mj} even if Taub-NUT metric has
special property such as the solutions of the metric are not
asymptotically flat (AF) but asymptotically locally flat
(ALF)~\cite{Hawking:1998jf,Hawking:1998ct}.
Hence, it is of interest
to investigate the corresponding situation in the case of warped
Taub-NUT/Bolt AdS black string.

It is well known that black string solution~\cite{Chamblin:1999by}
is suggested in order to investigate how
the physics of black holes is affected from the extra dimension with warped geometry
in Randall-Sundrum
(RS) brane world models~\cite{Randall:1999ee, Randall:1999vf}.
Such solution is noting but RS brane world models with appearing as a Schwarzchild
black hole on the brane.
In this case, it is found that the entropy is proportional
to the area of the event horizon on the brane and in the
bulk~\cite{Medved:2001zw,Jassal:2006sa}.
Therefore, an intriguing question is whether an area law is valid
for warped Taub-NUT AdS black string.

On the other hand, it has been recently suggested that considering $(d+1)$-dimensional AdS black holes,
the cosmological constant $\Lambda$ can be treated as the thermodynamic pressure $p$
\bear\label{pressure}
p=-\frac{1}{8\pi}\Lambda=\frac{(d-1)d}{16\pi l^2},
\eear
in units where $G=c=\hbar=k_B=1$.
Several series of relevant investigations have been performed
in this direction~\cite{Kastor:2009wy}$-$\cite{Lee:2015wua}.
Recently, it has been shown that there is a negative thermodynamic volume
in the Taub-NUT-AdS case~\cite{Johnson:2014xza} and found that there is
the first order phase transition from Taub-NUT-AdS to Taub-Bolt-AdS ~\cite{Johnson:2014pwa}.
The thermodynamic properties have been investigated extensively
in higher dimensional NUT/Bolt case~\cite{Lee:2014tma} and topological NUT/Bolt case~\cite{Lee:2015wua}.
Therefore, it would be interesting to be a similar discussion of the generalizations
in warped Taub-NUT/Bolt AdS black string. In this paper, we address these questions.

The paper is organized as follows: In the next section, we will yield
warped Taub-NUT AdS black string obtained by solving Einstein equation
with a negative cosmological constant in five-dimensional spacetime. In section 3,
we will calculate an entropy and
show how such entropy is expressed in terms of a cutoff parameter and a black hole's area.
Next, in the case of Taub-Bolt, we will explore the entropy and show it
is also proportional to being quadratically divergent in a cutoff parameter. In section 4,
when a cosmological constant is treated as a pressure,
we will explicitly calculate the thermodynamic quantities such
as the entropy, the enthalpy, the specific heat, the temperature,
the thermodynamic volume, and the Gibbs free energy. We will discuss their
extended thermodynamic properties. In the last section, we will give our conclusion.

\section{Warped Taub-NUT AdS Black String}
Warped Taub-NUT black string in five-dimensional AdS spacetime is given as
\bear\label{me1}
ds_5&=&a^2(z)\left[-f(r)(dt+2n\cos(\theta)d\phi)^2+\frac{dr^2}{f(r)}
+(r^2+n^2)(d\theta^2+\sin^2(\theta)d\phi^2)\right]\nonumber\\
&&+l_5^2dz^2,
\eear
which satisfies the Einstein equation with a negative cosmological constant
$\Lambda_{5}=-6/l_5$
\bear
G_{AB}-\Lambda_{5} g_{AB}=0.
\eear
Here, $l_5$ is the AdS radius, the warp factor $a(z)=\cosh(z)$, and
\bear
f(r)=\frac{r^2-n^2-2Mr+l_5^{-2}(r^4+6n^2r^2-3n^4)}{r^2+n^2}.\nonumber
\eear
Taking $\cosh(z)=1/\cos(\varphi)$, warped Taub-NUT black string in AdS$_5$
is rewritten as
\bear\label{redme1}
ds_5 &=&\frac{l_5^2}{\cos^2(\varphi)}\bigg[
\frac{1}{l_4^2}\left(-f(r)(dt+2n\cos(\theta)d\phi)^2+\frac{dr^2}{f(r)}
+(r^2+n^2)(d\theta^2+\sin^2(\theta)d\phi^2)\right)\nonumber\\
&&~~~~~~~~~~~~+d\varphi^2\bigg],
\eear
by introducing the relation between the cosmological parameter $l_5$ in the bulk
and the cosmological parameter $l_4$ on the blane
~\cite{Gregory:1993vy,Gregory:1994bj}
\bear\label{rel}
l_5=l_4\cos(\varphi)
\eear
The metric~(\ref{redme1}) on locally constant $\varphi=\varphi_0$ slice
is reduced to AdS Taub-NUT metric~\cite{Awad:2000gg} localized on a brane
\bear
ds_4=-f(r)(dt+2n\cos(\theta)d\phi)^2+\frac{dr^2}{f(r)}
+(r^2+n^2)(d\theta^2+\sin^2(\theta)d\phi^2).
\eear
We consider a scalar field $\Phi$ with mass $m$ propagating in
five-dimensional spacetime under the background (\ref{me1}), which is
described by five-dimensional Klein-Gordon equation as

\bear\label{KGe}
\frac{1}{\sqrt{-\det(g_{\mu\nu})}}\partial_{\mu}\left(\sqrt{-\det(g_{\mu\nu})}
\,g^{\mu\nu}\partial_{\nu}\Phi\right)-m^2\Phi=0
\eear
Then we set the wave function
\bear\label{KGe01}
\Phi(t,r,\theta,\phi,z)=\Psi(t,r,\theta,\phi)\chi(z),
\eear
which leads to
\bear
\frac{1}{a^4(z)l_5^2}\partial_z\bigg(a^4(z)\partial_z\chi(z)\bigg)
-(m^2-a^2(z)\mu^2)\chi(z)=0,
\eear
\bear\label{KGe02}
&&\left(\frac{4n^2\cos^2(\theta)}{(r^2+n^2)\sin^2(\theta)}-\frac{1}{f(r)}\right)
\partial_t^2\Phi-\frac{4n\cos(\theta)}{(r^2+n^2)\sin^2(\theta)}\partial_t\partial_{\phi}\Phi\nonumber\\
&&+\frac{1}{r^2+n^2}\partial_{r}\bigg((r^2+n^2)f(r)\partial_r\Phi\bigg)
+\frac{1}{(r^2+n^2)\sin(\theta)}\partial_{\theta}\bigg(\sin(\theta)\partial_{\theta}\Phi\bigg)\nonumber\\
&&+\frac{1}{(r^2+n^2)\sin^2(\theta)}\partial_{\phi}^{2}\Phi-\mu^2\Phi=0,
\eear
where $\mu$ can be interpreted as the effective mass on the brane
by introducing picture borrowed from RS brane world models~\cite{Randall:1999ee, Randall:1999vf}.
Since there are no explicit time dependent terms in Eq.(\ref{KGe02})
one may take the stationary solutions as the following
\bear\label{KGe01}
\Psi(t,r,\theta,\phi) = e^{-i\omega t}e^{i\alpha\phi}R(r)Q(\theta),
\eear
and substituting in Eq.(\ref{KGe02})
\bear\label{RedKGe01}
\frac{1}{\sin(\theta)}\partial_{\theta}\bigg(\sin(\theta)\partial_{\theta}Q(\theta)\bigg)
+\left[l(l+1)-\frac{\alpha^2+2\alpha(2\omega n \cos(\theta))
+4\omega^2n^2}{\sin^2(\theta)}\right]Q(\theta)=0,
\eear
\bear\label{RedKGe02}
\partial_{r}\bigg((r^2+n^2)f(r)\partial_{r}R(r)\bigg)
-\left[\left(\mu^2-\frac{\omega^2}{f(r)}\right)(r^2+n^2)+l(l+1)-4\omega^2n^2\right]R(r)=0,
\eear
where $l$ is the degree of spherical harmonic.
Using Wentzel-Kramers-Brillouin (WKB) approximation, the $z$-dependent wave number is given as
\bear\label{kz}
k_{z}^2 = l_5^2(m^2-a^2(z)\mu^2),
\eear
and the $r$-dependent wave number
\bear\label{kr}
k_{r}^2=\frac{h^2(r,l,\omega)}{f^2(r)},
\eear
with
\bear
h^2(r,l,\omega)=\left[\left(\frac{4n^2f(r)}{r^2+n^2}+1\right)\omega^2
-\left(\frac{l(l+1)}{r^2+n^2}+\mu^2\right)f(r)\right].
\eear

\section{Statistical Entropy \& Thermal Energy}
In this section, we will explicitly calculate the entropy through counting of the quantized modes for
the $z$-dependent wave number~(\ref{kz}) and the $r$-dependent wave number~(\ref{kr}).

The extra dimensional degeneracy factor $n_z$ is obtained
through the following the semi-classical quantization condition
\bear\label{nz}
\pi n_z &=&\int_{0}^{z_c}dzk_z(z,\mu),
\eear
and the radial degeneracy factor $n_r$
\bear\label{nr}
\pi n_r = \int_{r_h+\epsilon}^{D}drk_{r}(r,l,\omega).
\eear
From Eqs. (\ref{kz}) and (\ref{nz}), we obtain
\bear
\frac{dn_z}{d\mu} = \frac{ml_5}{\pi\mu}\gamma(\mu),
\eear
with
\bear
\gamma(\mu)={\bold E_2}\bigg(\sin^{-1}\big(a(z_c)\big),\frac{\mu^2}{m^2}\bigg)
-{\bold F_1}\bigg(\sin^{-1}\big(a(z_c)\big),\frac{\mu^2}{m^2}\bigg)
-{\bold E_1}\bigg(\frac{\mu^2}{m^2}\bigg)+{\bold K_1}\bigg(\frac{\mu^2}{m^2}\bigg),
\eear
where ${\bold E_1}$ is the complete elliptic integral, ${\bold E_2}$ the elliptic integral of the second kind,
${\bold F_1}$ the elliptic integral of the second kind,
and ${\bold K_1}$ the complete elliptic integral of the first kind.
Finally, in the WKB limit, the total number of quantum state with energy not exceeding $\omega$ is expressed as
\bear
g(\omega)=\int dg(\omega)=\int dl (2l+1)\int d\mu \left(\frac{dn_z}{d\mu}\right)\int \pi dn_r.
\eear
The free energy $F$ of the scalar field at inverse temperature $\beta$ yields
\bear
\pi\beta F &=& -\int_{0}^{\infty}d\omega\frac{\beta g(\omega)}{e^{\beta\omega}-1}\nonumber\\
&=&-\beta\int_{0}^{\infty}\frac{d\omega}{e^{\beta\omega}-1}\int dl (2l+1)
\int \frac{ml_5}{\pi}\frac{\gamma(\mu)d\mu}{\mu}\nonumber\\
&&\times \int_{r_h+\epsilon}^{L}\frac{dr}{f(r)}\sqrt{\left(\frac{4n^2f(r)}{r^2+n^2}+1\right)\omega^2
-\left(\frac{l(l+1)}{r^2+n^2}+\mu^2\right)f(r)}.
\eear
Here, the reality condition of the free energy leads to the following limits for the remaining integrals:
\bear
0\leq l \leq \frac{1}{2}\left[-1+\sqrt{1-4(n^2+r^2)\left\{\left(\frac{4n^2 f(r)}{r^2+n^2}+1\right)
\frac{\omega^2}{f(r)}-\mu^2\right\}}\right],
\eear
and
\bear
m\leq \mu \leq \sqrt{\frac{4n^2 f(r)}{r^2+n^2}+1}\frac{\omega}{\sqrt{f(r)}}.
\eear
On the integrating over $l$, the free energy $F$ is given as
\bear
F&=&-\frac{2ml_5}{3\pi}\int_{0}^{\infty}\frac{d\omega}{e^{\beta\omega}-1}\int \frac{\gamma(\mu)d\mu}{\mu}\nonumber\\
&&\times
\int_{r_h+\epsilon}^{L} dr\left(\frac{r^{2}+n^2}{f^2(r)}\right)
\left[\left(\frac{4n^2f(r)}{r^2+n^2}+1\right)\omega^2
-\mu^2 f(r)\right]^{\frac{3}{2}}.
\eear
The main contributions to the free energy, on carrying out the integrals over $\mu$, are
\bear
F\approx -\rho(z_c)\frac{2ml_5}{3\pi}\int_{0}^{\infty}d\omega\frac{\omega^3}{e^{\beta\omega}-1}
\int_{r_h+\epsilon}^{L} dr\left(\frac{r^{2}+n^2}{f^2(r)}\right)
\eear
with
\bear\label{rho}
\rho(z_c)=a(z_c)-1,
\eear
which in the approximation of $L\gg r_h$ leads to
\bear\label{freeenergy}
F \thickapprox \left[-\frac{2\pi^3m}{45\epsilon\beta^4}
\frac{r_{h}^{2}(r_{h}^{2}+n^2)}{\bigg(3(r_{h}^{2}+n^2)l_{5}^{-2}+1\bigg)^2}\right]\rho(z_c)l_5,
\eear
where the followings are used:
\bear
f(r)\approx (r-r_h)\left.\frac{df(r)}{dr}\right|_{r=r_h}=(r-r_h)\bigg(3(r_h^2+n^2)l_5^{-2}+1\bigg),
\eear
and
\bear
\int_{0}^{\infty}d\omega\frac{\omega^3}{e^{\beta\omega}-1}=\frac{\pi^4}{15\beta^4}.
\eear
The entropy is given as
\bear\label{etropy01}
S=\beta^2\left(\frac{\partial F}{\partial \beta}\right)
=\left[\frac{8\pi^3m}{45\epsilon\beta^3}\frac{r_h^2(r_h^2+n^2)}
{\bigg(3(r_h^2+n^2)l_{5}^{-2}+1\bigg)^2}\right]\rho(z_c)l_5.
\eear
The cutoff near the horizon is referred to as the brick wall.
The physical distance between the brick wall and the horizon is
\bear
\int_{r_h}^{r_h+h}ds=\int_{r_h}^{r_h+\epsilon}\frac{dr}{\sqrt{f(r)}},
\eear
which leads to
\bear\label{eps}
\epsilon=\left[\frac{3(r_h^2+n^2)l_{5}^{-2}+1}{4r_h}\right]h^2.
\eear
In terms of this covariant cutoff parameter $h$,
the ultraviolet divergent part of the scalar field entropy is given by
\bear\label{etropy03}
S=\left[\frac{32\pi^3m}{45h^2\beta^3}\frac{r_h^3(r_h^2+n^2)}
{\bigg(3(r_h^2+n^2)l_{5}^{-2}+1\bigg)^3}\right]\rho(z_c)l_5,
\eear
which is written in terms of the parameter $M$
\bear\label{etropy04}
S=\frac{4\pi^3m}{45h^2\beta^3M^2}\left[
\frac{\bigg(r_h^2-n^2+l_5^{-2}(r_h^4+6n^2r_h^2-3n^4)\bigg)^3}
{(r_h^2+n^2)^2\bigg((r_h^2+n^2)^{-1}+3l_5^{-2}\bigg)^3}\right]\rho(z_c)l_5.
\eear
When taking $M=n+4(2+\sqrt{2})n^3l_5^{-2}$, $r_h$ is evaluated as $(\sqrt{2}+1)n$.
Imposing the inverse temperature $\beta=8\pi n$, the entropy in warped Tabu-NUT AdS spacetime
is obtained as
\bear\label{etropy05}
S_{\rm AdS,NUT}=\left[\frac{0.13n^2m}{(20.49n^2l_5^{-2}+1)^3h^2}\right]\rho(z_c)l_5.
\eear
Taking $M=((41\sqrt{41}+365)n^2l_5^{-2}+40)n/32$ in warped Tabu-Bolt AdS spacetime, $r_h$ is
given by $(\sqrt{41}+5)(n/4)$, and the entropy
\bear\label{etropy06}
S_{\rm AdS,Bolt}=\left[\frac{0.29n^2m}{(27.38n^2l_5^{-2}+1)^3h^2}\right]\rho(z_c)l_5.
\eear
They reveal that
the entropy in warped Tabu-Taub-NUT/Bolt AdS spacetime has an explicit dependence on
the inversely square of the cutoff parameter $h$.

Since the area of the horizon ${\cal A}_h$ on the brane located at $z=z_c$ is
\bear
{\cal A}_h=\int_{0}^{2\pi}\int_{0}^{\pi}\sqrt{g_{\theta\theta}g_{\phi\phi}}d\theta d\phi=4\pi a^2(z_c)(r_h^2+n^2),
\eear
the above entropy (\ref{etropy05}) is written as
\bear\label{enads}
S_{\rm AdS,NUT}=\left[\left(\frac{\rho(z_c)}{16(\sqrt{2}+1)\pi a^2(z_c)}\right)
\left(\frac{0.13m}{(20.49n^2l_5^{-2}+1)^3}\right)\left(\frac{l_5}{h^2}\right)\right]{\cal A}_{h},
\eear
which leads to
\bear
S_{\rm AdS,NUT}=\left[\left\{\left(\frac{\rho(z_c)}{16(\sqrt{2}+1)\pi a^2(z_c)}\right)
\left(\frac{0.13ml_4 \cos(\varphi)}{(\frac{20.49n^2l_4^{-2}}{\cos^{2}(\varphi)}+1)^3}\right)\right\}
\left(\frac{1}
{h^2}\right)\right]{\cal A}_{h},
\eear
by using the relation (\ref{rel}).
Evidently, the entropy on the brane located at $z=z_c$ satisfies
the black hole area law
since the curly-brackets quantity is much more smaller than the cutoff parameter $h$
if we take $h$ to be of the order of Planck length.

Introducing the invariant length of the black string along the $z$-direction
\bear
\int_{0}^{z_c}dz\sqrt{g_{zz}}=z_c l_5  \equiv {\cal R},
\eear
from the entropy (\ref{enads}) we have in the limit of enough large ${\cal R}$
\bear
S_{\rm AdS,NUT}\thickapprox\left[\left(\frac{0.13\rho(z_c)m}{16(\sqrt{2}+1)\pi z_c a^2(z_c)}\right)
\left(\frac{1}{h^2}\right)\right]{\cal A}_{\rm BS},
\eear
where the area of the black string horizon $\cal A_{\rm BS}$  in the bulk is given as
\bear
{\cal A}_{\rm BS}={\cal A}_h\times {\cal R}.
\eear
It is also shown that the entropy of warped Taub-NUT AdS black string in the bulk
holds for an area law from assumption of $h$ with the order of Planck length.

One the other hand, employing thermal relation ${\cal E}=\partial_{\beta} (\beta F)$,
the thermal energy  ${\cal E}$ is given as
\bear
{\cal E}=\left[\frac{2\pi^3m}{15\epsilon\beta^4}
\frac{r_{h}^{2}(r_{h}^{2}+n^2)}{\bigg(3(r_{h}^{2}+n^2)l_{5}^{-2}+1\bigg)^2}\right]\rho(z_c)l_5.
\eear
Taking $r_h=(\sqrt{2}+1)n$ and $\beta=8\pi n$ and substituting in (\ref{eps}),
we obtain the thermal energy in the case of Taub-NUT  ${\cal E_{\rm NUT}}$
\bear
{\cal E_{\rm NUT}}=\left[\frac{(17\sqrt{2}+24) m n}{3840
\bigg(6(2+\sqrt{2})n^2l_5^{-2}+1)\bigg)^3}\right]\rho(z_c)l_5,
\eear
and the thermal energy in the case of Taub-Bolt ${\cal E_{\rm Bolt}}$
\bear
{\cal E_{\rm Bolt}}=\left[\frac{(1057\sqrt{41}+6765) m n}{960
\bigg(3(5\sqrt{41}+41)n^2l_5^{-2}+8)\bigg)^3}\right]\rho(z_c)l_5,
\eear
by taking  $r_h=(\sqrt{41}+5)(n/4)$.
It is shown that both thermal energies depend on the bulk parameter $z_c$ along the $z$-direction
and diverge much more slowly as $n$ grows up.

\section{The  Extended Thermodynamic Properties}
Employing Wick rotation the time ($t\rightarrow it$) and the NUT charge ($n\rightarrow iN$),
the Euclidean section for warped Taub-NUT AdS black string is obtained as
\bear\label{eI}
ds_5&=&a^2(z)\left[g(r)(dt+2N\cos(\theta)d\phi)^2+\frac{dr^2}{g(r)}
+(r^2-N^2)(d\theta^2+\sin^2(\theta)d\phi^2)\right]\nonumber\\
&&+l_5^2dz^2,
\eear
with \bear
g(r)=\frac{r^2+N^2-2Mr+l_5^{-2}(r^4-6N^2r^2-3N^4)}{r^2-N^2}.\nonumber
\eear

\begin{figure}[!htbp]
\begin{center}
{\includegraphics[width=12cm]{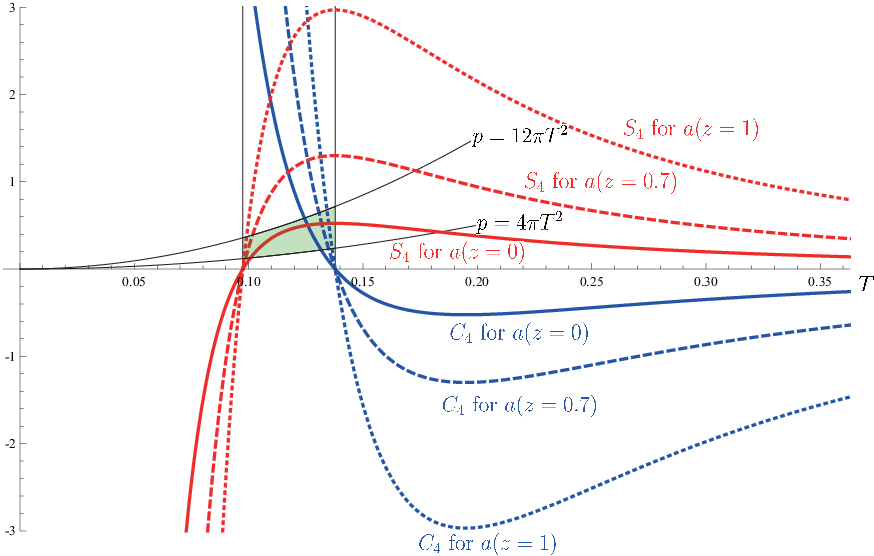}}
\end{center}
\vspace{-0.6cm}
\caption{{\footnotesize Plot of the entropy $S_4$ (red dotted curve for $a(z=1)$,
red dashed curve for $a(z=0.7)$, and red solid curve for $a(z=0)$, respectively),
specific heat $C_4$ (blue dotted curve for $a(z=1)$,
blue dashed curve for $a(z=0.7)$, and blue solid curve for $a(z=0)$, respectively)
and pressure $p$ (the upper black solid curve for $p=12\pi T^2$, and
the lower black solid curve for $p=4\pi T^2$, respectively)
as a function of the temperature $T$ in five dimensions.}}
\label{figI}
\end{figure}

Using calculation of the period of the Euclidean section (\ref{eI}),
one can get the inverse temperature $\beta$
\bear\label{EIT}
\beta=8\pi N.
\eear
Employing counter term subtraction method, the regularized action $I_{\rm NUT}$ is given as
\bear
I_{\rm NUT}=\frac{4a^4(z)\pi (l_5^2-2N^2)}{l_5}.
\eear
Using the Gibbs-Duhem relation $S=\beta M-I$, the entropy $S_{\rm NUT}$ is
\bear\label{SN}
S_{\rm NUT}=\frac{a^4(z)\sqrt{3}(4\pi T^2-p)}{128\sqrt{2}\pi^{\frac{5}{2}}T^4\sqrt{p}},
\eear
and by thermal relation $C=-\beta \partial_{\beta}S$, the specific heat $C_{\rm NUT}$ is given as
\bear\label{CN}
C_{\rm NUT}=\frac{a^4(z)\sqrt{3}(p-2\pi T^2)}{32\sqrt{2}\pi^{\frac{5}{2}}T^4\sqrt{p}},
\eear
and by thermal relation $H=U+pV$, the enthalpy $H_{\rm NUT}$ is
\bear
H_{\rm NUT}=\frac{a^4(z)(6\pi T^2-p)}{32\sqrt{6}\pi^{\frac{5}{2}}T^3\sqrt{p}},
\eear
and the thermodynamic volume $V_{\rm NUT}$
\bear
\partial_{p} H_{\rm NUT}\bigg|_{S_{\rm NUT}}
=V_{\rm NUT}=-\frac{a^4(z)(36\pi T^2+7p)}{235\sqrt{6}\pi^{\frac{5}{2}}p^{\frac{3}{2}}T^3},
\eear
which shows that the thermodynamic volume of warped Taub-NUT AdS black string is negative
like results of Taub-NUT AdS solutions~\cite{Johnson:2014xza,Lee:2014tma,Lee:2015wua}.
The above thermodynamic quantities satisfy the generalized Smarr formula
~\cite{Kastor:2009wy,Cvetic:2010jb,Caldarelli:1999xj}
\bear
H_{\rm NUT}-2TS_{\rm NUT}+2pV_{\rm NUT}=0,
\eear
which is precisely matched with that of five-dimensional
Taub-NUT solution with negative cosmological constant~\cite{Lee:2014tma,Lee:2015wua}.
Using an thermal relation $U=H-pV$, the internal energy $U_{\rm NUT}$ is obtained as
\bear
U_{\rm NUT}=\frac{a^4(z)(84\pi T^2-p)}{256\sqrt{6}\pi^{\frac{5}{2}}T^3\sqrt{p}}
\eear

\begin{figure}[!htbp]
\begin{center}
{\includegraphics[width=12cm]{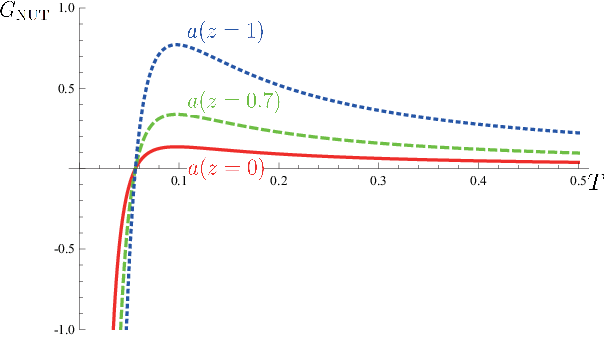}}
\end{center}
\vspace{-0.6cm}
\caption{{\footnotesize Plot of the five-dimensional Gibbs free energy $G$
as a function of temperature $T$ for $p=1$ (blue dotted curve for $a(z = 1)$, green dashed
curve for $a(z = 0.7)$, and red solid curve for $a(z = 0)$, respectively).}}
\label{figII}
\end{figure}

Finally, employing the Legendre transform of enthalpy $G=H-TS$, the Gibbs free energy is given as
\bear\label{GN}
G_{\rm NUT}=\frac{a^4(z)(12\pi T^2-p)}{128\sqrt{6}\pi^{\frac{5}{2}}T^3\sqrt{p}}
\eear
Requiring the specific heat are positive, one can obtain the following thermally stable range
of $T$ for warped Taub-NUT AdS black string
\bear
\sqrt{\frac{p}{4\pi}}<T<\sqrt{\frac{p}{2\pi}}.
\eear
However, since the Gibbs free energy $G_{\rm NUT}$ is positive,
warped Taub-NUT AdS black string in some areas of this region is still unstable for $p<4\pi T^2$
or for $p>12\pi T^2$. Finally, requiring the Gibbs free energy $G_{\rm NUT}$  are negative
$4\pi T^2<p<12\pi T^2$, warped Taub-NUT AdS black string becomes a thermally stable
(such black string in the light-green shaded areas is thermally stable in Fig. 1).
A similar result is obtained in \cite{Lee:2015wua}.
As the warp factor $a(z)=\cosh(z)$ grows up, the entropy $S_{\rm NUT}$ (\ref{SN}),
the specific heat $C_{\rm NUT}$ (\ref{CN}), and the Gibbs free energy $G_{\rm NUT}$ (\ref{GN})
increase (as you see in Fig. 1 and Fig. 2).

Let us consider the Bolt case ($r=r_{\rm B}>N$). Using parallel way as in the case of the NUT case,
The inverse of the temperature $\beta$, the action $I_{\rm Bolt}$,
and the enthalpy $H_{\rm Bolt}$ are respectively
\bear
\beta=\left.\frac{4\pi}{f'(r)}\right|_{r=r_{\rm B}}
=\frac{4\pi l^2 r_{\rm B}}{kl^2+(2u+1)(r_{\rm B}-N^2)},
\eear
\bear
I_{\rm Bolt}=\frac{2a^4(z)\pi N \{-3N^4-r_{\rm B}^4+(N^4+r_{\rm B}^2)l_5^2\}}{r_{\rm B}l_5^2},
\eear
with the Bolt radius $r_{\rm B}$
\bear
r_{\rm B,\pm}=\frac{l_5^2\pm\sqrt{l_5^4-48l_5^2N^2+144N^4}}{12N},
\eear
where requiring the discriminant of the square root in the Bolt radius $r_{\rm B,\pm}$ is positive,
the maximum magnitude of the NUT charge $N_{\rm max}$ is obtained as
\bear
N\leq \frac{l_5}{2\sqrt{3(2+\sqrt{3})}}=N_{\rm max},
\eear
The enthalpy $H_{\rm Bolt}$, the entropy $S_{\rm Bolt}$, and thermodynamic volume $V_{\rm Bolt}$
for the Bolt solution are given as
\bear
H_{\rm Bolt}=\frac{a^4(z)\{-3p+24\pi(-16\pi r_{\rm B}^2p+1)T^2+512\pi^3 r_{\rm B}^2(8\pi r_{\rm B}^2p+3)T^4\}}
{2048\sqrt{6}\pi^{\frac{7}{2}}r_{\rm B}T^4\sqrt{p}},
\eear
\bear
S_{\rm Bolt}=\frac{a^4(z)\sqrt{3\pi}\{64\pi^2(\frac{4p}{p-8\pi(8\pi r_{\rm B}^2p+1)T^2}+1)r_{\rm B}^2T^2+1\}}
{128\pi^2T^2\sqrt{2p}},
\eear
\bear
V_{\rm Bolt}=\frac{a^4(z)\sqrt{2\pi r_{\rm B}(64\pi^2 r_{\rm B}^2T^2-3)}}{64\pi^2 T^2 \sqrt{3p}}.
\eear
Finally, the specific heat $C_{\rm Bolt}$, the internal energy $U_{\rm Bolt}$, and
the Gibbs free energy $G_{\rm Bolt}$ yield respectively
\bear
C_{\rm Bolt}=-\frac{a^4(z)\sqrt{3\pi}T^2}{8\sqrt{2}p^{\frac{5}{2}}}
\pm\frac{a^4(z)\sqrt{3}(p^4-6\pi p^3 T^2+2\pi^2 p^2 T^4-8\pi^3 p T^6+8\pi^4 T^8)}
{32\pi^{\frac{5}{2}}p^{\frac{5}{2}}T^4\sqrt{2p^2-16\pi p T^2+8\pi^2 T^4}},
\eear
\bear
U_{\rm Bolt}=\frac{-p+8\pi(1-8\pi r_{\rm B}^2p)T^2+512\pi^3(r_{\rm B}+2p)r_{\rm B}T^4}{1024\pi^3r_{\rm B}T^4},
\eear
\bear
G_{\rm Bolt}=\frac{a^4(z)(1-32\pi^2T^2+2048\pi^4T^4)}{256\pi^{\frac{7}{2}}T^3\sqrt{2p}}.
\eear

\begin{figure}[!htbp]
\begin{center}
{\includegraphics[width=12cm]{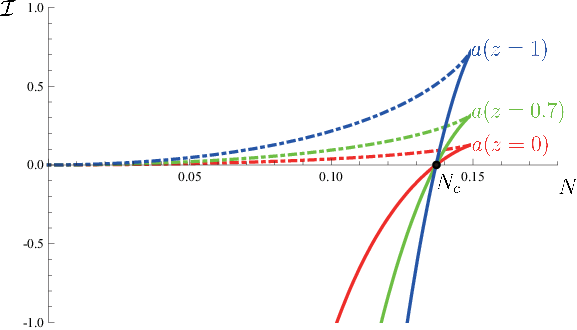}}
\end{center}
\vspace{-0.6cm}
\caption{{\footnotesize Plot of the action difference as a function of $N$
(blue curve for a($z = 1$), green
curve for a($z = 0.7$), and red curve for a($z = 0$), respectively, when $p=3/8\pi$).}}
\label{figIII}
\end{figure}

From now on, considering the action difference of warped Taub-NUT AdS black string
and warped Taub-Bolt AdS black string, we investigate their instability.

The action difference, $I_{\rm Bolt}-I_{\rm NUT}={\cal I}$, is obtained as
\bear\label{AD}
{\cal I}=\frac{\sqrt{\pi} a^4(z)N(N-r_{\rm B})^2\{3-8\pi(3N^2+2Nr_{\rm B}+r_{\rm B}^2)p\}}
{r_{\rm B}\sqrt{6p}}.
\eear
As shown in Fig. 3, for the pressure $p$ as fixed parameter (isobaric process),
the action difference ${\cal I}$ (\ref{AD}) becomes negative when $N$ increases beyond a critical
NUT charge $N_c$. This implies that the first order phase transition
from warped Taub-NUT-AdS black string to warped Taub-Bolt-AdS black string occurs at $N_c$.
Then, after getting $N_c$ through solving the action difference ${\cal I}=0$ and
substituting into the inverse of the temperature (\ref{EIT}), the critical temperature $T_c$
is given as
\bear\label{CT}
\frac{(\sqrt{5}+\sqrt{2})\sqrt{p}}{\sqrt{6\pi}}.
\eear
As shown in Fig. 3, the action difference ${\cal I}$ (\ref{AD}) increases
as the warp factor $a(z)=\cosh(z)$ grows up.

\section{Conclusion}
Previous studies of quantized modes of scalar fields in Taub-NUT
background~\cite{Ghosh:2002mj} have indicated that the cutoff
parameter $h$ is determined by matching the statistical entropy with
the gravitational entropy even if it is changed by the relative
values of the parameters $M$ and $n$.

Here we found the similar
results for warped Taub-NUT AdS black string. Our evaluations of
quantized modes of scalar fields in various warped black strings
background were shown that the cutoff parameter has the Planck scale
only by appropriate choosing the size of the AdS radius $l_5$, in
contrast to the results in as~\cite{Randall:1999ee, Randall:1999vf}.
We examined the correction to the statistical entropy
due to the extra dimension with warped geometry and found the
statistical entropy is still quadratically divergent in the
cutoff parameter as in Taub-NUT metric~\cite{Ghosh:2002mj}.

Next, we showed that the entropy of warped Taub-NUT AdS black
satisfies an area law in the bulk as well as on the brane.
Furthermore, we investigated the thermal energy
in warped Taub-NUT AdS/Bolt black string and found that
both thermal energies depend on the bulk parameter $z_c$ along
the $z$-direction and diverge much more slowly as $n$ grows up.

Finally, we explicitly calculated their thermal quantities the context of the extended
thermodynamics. It was found out that there existed the first order phase transition
from warped Taub-NUT-AdS black string to warped Taub-Bolt-AdS black string.
Furthermore, we obtained thermodynamically
stable region as of a function of the temperature $T$ for NUT case.
We also investigated a proportional behavior of the thermodynamic quantities
with respect to the warp factor
and found that the entropy, the specific heat, the Gibbs free energy
and the action difference are proportional to the warp factor.

Recently, an instability issue was covered for all topological spacetime
through investigations of quasinormal modes of warped black strings
with nontrivial topologies in AdS$_{5}$~\cite{Yin:2010ix}. In this
context it would be of interest to study the stability of our warped
black string solution.
It was shown that four-dimensional BTZ black string~\cite{Liu:2008ds}
still holds for the Gubser-Mitra conjecture~\cite{Gubser:2000mm},
correlating the classical and thermodynamic instabilities of black strings.
Intriguing issue would be to accomplish to test this conjecture
for warped Taub-NUT AdS black string.


\end{document}